\documentclass[12pt]{article}
\usepackage{fullpage}
\usepackage{natbib}
\usepackage{amsmath,amssymb}
\usepackage{graphicx}
\newcommand{\bSig}{\boldsymbol{\Sigma}}
\newcommand{\bK}{\boldsymbol{K}}
\newcommand{\bZ}{\boldsymbol{Z}}
\newcommand{\bY}{\boldsymbol{Y}}
\newcommand{\bX}{\boldsymbol{X}}
\newcommand{\bA}{\boldsymbol{A}}
\newcommand{\bB}{\boldsymbol{B}}
\newcommand{\bD}{\boldsymbol{D}}
\newcommand{\bU}{\boldsymbol{U}}
\newcommand{\bS}{\boldsymbol{S}}
\newcommand{\bs}[1]{\boldsymbol{#1}}
\newcommand{\mc}[1]{\mathcal{#1}}
\newcommand{\inner}[1]{\langle #1 \rangle}
\newcommand{\Pos}{\bf{P}}
\newcommand{\bPhi}{\bs{\Phi}}
\newcommand{\bPsi}{\bs{\Psi}}

\begin{document}

\title{A Multivariate Graphical Stochastic Volatility Model}
\author{Yuan Cheng and Alex Lenkoski\footnote{\noindent \textit{Corresponding author address:} Alex Lenkoski, Institute of Applied Mathematics, Heidelberg University, Im Neuenheimer Feld 294, 69120 Heidelberg, Germany
\newline{E-mail: alex.lenkoski@uni-heidelberg.de}}\\
\textit{Heidelberg University, Germany}}
\maketitle
\begin{abstract}
\noindent
The Gaussian Graphical Model (GGM) is a popular tool for incorporating sparsity into joint multivariate distributions. The G-Wishart distribution, a conjugate prior for precision matrices satisfying general GGM constraints, has now been in existence for over a decade.  However, due to the lack of a direct sampler, its use has been limited in hierarchical Bayesian contexts, relegating mixing over the class of GGMs mostly to situations involving standard Gaussian likelihoods.  Recent work, however, has developed methods that couple model and parameter moves, first through reversible jump methods and later by direct evaluation of conditional Bayes factors and subsequent resampling.  Further, methods for avoiding prior normalizing constant calculations--a serious bottleneck and source of numerical instability--have been proposed.  We review and clarify these developments and then propose a new methodology for GGM comparison that blends many recent themes.  Theoretical developments and computational timing experiments reveal an algorithm that has limited computational demands and dramatically improves on computing times of existing methods.  We conclude by developing a parsimonious multivariate stochastic volatility model that embeds GGM uncertainty in a larger hierarchical framework. The method is shown to be capable of adapting to the extreme swings in market volatility experienced in 2008 after the collapse of Lehman Brothers, offering considerable improvement in posterior predictive distribution calibration.

\end{abstract}
\newpage
\section{Introduction}
\indent The Gaussian graphical model (GGM) has received widespread consideration \citep[see][]{jones_et_2005} and estimators obeying graphical constraints in standard Gaussian sampling were proposed as early as \citet{dempster_1972}.  Initial incorporation of GGMs in Bayesian estimation largely focus on decomposable graphs \citep{dawid_lauritzen_1993}, since prior distributions factorize into products of Wishart distributions.  \citet{roverato_2002} proposes a generalized extension of the Hyper-Inverse Wishart distribution for covariance matrices $\bSig$ over arbitrary graphs. \citet{atay-kayis_massam_2005} turn this into a prior specified for precision matrices $\bK$ and outline a Monte Carlo (MC) method that enables pairwise model comparisons.  Following \citet{letac_massam_2007} and \citet{rajaratnam_et_2008}, \citet{lenkoski_dobra_2011} term this distribution the G-Wishart, and propose computational improvements to direct model comparison and model search.\\
\indent A number of samplers for precision matrices under a G-Wishart distribution have been proposed.  These involve either block Gibbs sampling \citep{piccioni_2000}, Metropolis-Hastings (MH) moves \citep{mitsakakis_et_2011,dobra_lenkoski_2011, dobra_et_2011}, or rejection sampling \citep{wang_carvalho_2010}.  \citet{dobra_et_2011} shows that the rejection sampler of \citet{wang_carvalho_2010} suffers from extremely low acceptance probabilities in even moderate dimensions.  \citet{wang_li_2012} conclusively show that block Gibbs sampling is both computationally more efficient and exhibits considerably less autocorrelation than the MH methods.\\
\indent The block Gibbs sampler provides a Markov chain Monte Carlo (MCMC) sample.  When the likelihood assumes standard Gaussian sampling, determining posterior expectations of $\bK$ can technically be performed as in \citet{lenkoski_dobra_2011}, whereby model probabilities are first directly assessed via stochastic search, and model averaged samples are then collected using block Gibbs over each model.  However, when the GGM is specified over latent data in a hierarchical Bayesian framework, such an approach is no longer valid.  This is due to the use of the matrix $\bK$ in updating other hyperparameters as well as its involvement in updating the latent Gaussian factors.\\
\indent \citet{dobra_lenkoski_2011} propose a reversible jump MCMC (which for brevity we refer to as RJ) method \citep{green_1995} that simultaneously updates the GGM and its associated $\bK$, and embed the GGM in a semiparametric Gaussian copula.  \citet{dobra_et_2011} expand the RJ algorithm and show how GGMs may be used to model dependent random effects in a generalized linear model context, focusing on lattice data.  \citet{wang_li_2012} utilize conditional properties of G-Wishart variates that enables model moves through calculation of a conditional Bayes factor (CBF) \citep{dickey_gunel_1978} and subsequently update $\bK$ through direct Gibbs sampling.  \citet{wang_li_2012} also explore the use of a double MH algorithm \citep{liang_2010} to avoid the computationally expensive and numerically unstable MC approximation of normalizing constants proposed by \citet{atay-kayis_massam_2005}.\\
\indent We investigate an alternative method for simultaneously updating the GGM and associated $\bK$ in hierarchical Bayesian settings.  Our method is built on the framework outlined in \citet{wang_li_2012}, but blends several of the developments above to yield an algorithm with considerably less computational cost.  We show that use of a CBF on the Cholesky decomposition of a permuted version of $\bK$, originally proposed in an early version of \citet{wang_li_2012}, enables fast model moves; use of methods for sparse Cholesky decompositions \citep{rue_2001} further reduce computational overhead.  We then show that while both \citet{dobra_et_2011} and \citet{wang_li_2012} indicate the random walk MH sampler of \citet{mitsakakis_et_2011} is not suitable for posterior sampling, it is especially useful in the context of the double MH algorithm.  We are therefore able to specify a model move algorithm which requires little computational effort and exhibits no numerical instability.\\
\indent Simulation experiments compare our new algorithm to the algorithm of \citet{wang_li_2012} (which we refer to as WL). Both methods perform equally well at determining posterior distributions. However, we show that while the WL approach is theoretically appealing, it suffers significant computational overhead on account of many matrix inversions.  By contrast, our new approach exhibits a dramatic improvement in computation time.\\
\indent We conclude with an example of how GGMs may be embedded in hierarchical Bayesian models.  GGMs have been shown to yield parsimonious joint distributions useful in financial applications \citep{carvalho_west_2007, rodriguez_et_2011}.  However, existing studies have largely ignored heteroskedasticity in financial data and relied on datasets taken over periods with relatively little financial turmoil.  To address this, we propose a parsimonious multivariate stochastic volatility model that incorporates GGM uncertainty. We then model stock returns for $20$ assets during the period surrounding the financial crash of 2008.  We show that in the periods leading up to the crash, and 9 months after the most turbulent period, this method yields no improvement over an approach that does not model heteroskedasticity.  However, during the period of heightened volatility, our new model is able to adapt quickly and yields considerably more calibrated predictive distributions.\\
\indent The article is organized as follows.  In Section~\ref{sec:gwish} we review the G-Wishart distribution, establish results necessary for CBF calculations and describe the block Gibbs sampler.  Section~\ref{sec:simulation} conducts a simulation study showing the computational advantage gained by our new algorithm. In Section~\ref{sec:vol} we describe our multivariate graphical stochastic volatility model and give results over the data mentioned above.  We conclude in Section~\ref{sec:conclude}.
\section{The G-Wishart Distribution}\label{sec:gwish}
\subsection{Review of Basic G-Wishart Properties}
\indent Suppose that we collect data $\mc{D} = \{\bZ^{(1)}, \dots, \bZ^{(n)}\}$ such that $\bZ^{(j)}\sim \mc{N}_p(0,\bK^{-1})$ independently for $j \in \{1,\dots,n\}$, where $\bK\in\Pos$, the space of $p\times p$ positive definite matrices.  This sample has likelihood
$$
pr(\mc{D}|\bK) = (2\pi)^{-np/2}|\bK|^{n/2} \exp\left(-\frac{1}{2}\inner{\bK,\bU}\right),
$$
where $\inner{A,B} = tr(A'B)$ denotes the trace inner product and $\bU = \sum_{i = 1}^n \bZ^{(i)}\bZ^{(i)'}$.\\
\indent Further suppose that $G = (V,E)$ is a GGM where $V = \{1,\dots,p\}$ and $E \subset V\times V$.  We will slightly abuse notation throughout, by writing $(i,j) \in G$ to indicate that the edge $(i,j)$ is in the edge set $E$.  Associated with $G$ is a subspace $\Pos_G \subset \Pos$ such that $\bK\in \Pos_G$ implies that $\bK\in\Pos$ and $K_{ij} = 0$ whenever $(i,j) \not\in G$.  The G-Wishart distribution \citep{roverato_2002,atay-kayis_massam_2005} $\mc{W}_G(\delta,\bD)$ assigns prior probability to $\bK\in\Pos_G$ as
$$
pr(\bK|\delta,\bD,G) = \frac{1}{I_G(\delta,\bD)}|\bK|^{(\delta - 2)/2}\left(-\frac{1}{2}\inner{\bK,\bD}\right)\bs{1}_{\bK\in\Pos_G}.
$$
The normalizing constant $I_G$ is in general not known to have an explicit form, and \citet{atay-kayis_massam_2005} propose an MC approximation for this factor. Furthermore, the G-Wishart is conjugate and thus $pr(\bK|\mc{D}, G) = \mc{W}_G(\delta + n, \bD^*)$ where $\bD^{*} = \bD + \bU$.\\
\indent Let $\bPhi$ be the upper triangular matrix such that $\bPhi'\bPhi = \bK$, the Cholesky decomposition.  \citet{rue_2001} notes that we may associate with $G$ another graph $F$, called the \emph{fill-in} graph, such that $G\subset F$,  $\Phi_{ij} = 0$ when $(i,j)\notin F$ and
\begin{equation}\label{eq:completephi}
\Phi_{ij} = -\frac{1}{\Phi_{ii}}\sum_{l = 1}^{i}\Phi_{li}\Phi_{lj}
\end{equation}
when $i < j$ and $(i,j)\in F\setminus G$.  \citet{rue_2001} outlines a straightforward method for constructing a graph $F$ from $G$ and explains how use of node reordering software can minimize fill-in.\\
\indent \citet{roverato_2002} shows that if $K\sim \mc{W}_G(\delta, \bD)$ then
\begin{equation}\label{eq:phiwishart}
pr(\Phi|\delta,\bD,G) = \prod_{i = 1}^{p} \Phi_{ii}^{\delta + \nu_i^G - 1}\exp\left(-\frac{1}{2}\inner{\bPhi'\bPhi,\bD}\right),
\end{equation}
where $\nu_i^G$ is number of nodes in $\{i + 1, \dots, p\}$ that are connected to node $i$ in $G$.  We especially note that if $\bK\sim \mc{W}_G(\delta, \mathbb{I}_p)$, then
\begin{equation}\label{eq:phiwishartiden}
pr(\bPhi|\delta,\mathbb{I}_p,G) = \exp\left(-\frac{1}{2}\sum_{(i,j)\in F} \Phi_{ij}^2\right)\prod_{i = 1}^{p} \Phi_{ii}^{\delta + \nu_i^G - 1}\exp\left(-\frac{1}{2} \Phi_{ii}^2\right).
\end{equation}
\subsection{Sampling Methods}\label{sec:gwish_samplers}
\indent We review two MCMC methods for approximate sampling from a $\mc{W}_G(\delta,\bD)$.  See \citet{dobra_et_2011} and \citet{wang_li_2012} for more detailed reviews of the many methods that have been proposed.\\
\indent Let $\mc{C}$ denote the cliques of $G$.  In the following, we consider a clique to be a maximally complete subgraph, though \citet{wang_li_2012} note that this can be relaxed to any complete subgraph. \citet{piccioni_2000} shows that for any $C\in\mc{C}$,
\begin{equation}\label{eq:bgibbs_update}
K_C - K_{C,V\setminus C}K_{V\setminus C}^{-1}K_{V\setminus C,C} \sim \mc{W}(\delta, D_C),
\end{equation}
where $\mc{W}$ denotes a standard Wishart variate.  The expression (\ref{eq:bgibbs_update}) thereby gives the full conditionals of $\mc{W}_G(\delta,\bD)$.  The block Gibbs sampler thus cycles through $\mc{C}$, updating each component using (\ref{eq:bgibbs_update}).  \citet{wang_li_2012} convincingly show that for posterior inference of $\mc{W}_{G}(\delta + n, \bD^{*})$ the block Gibbs sampler outperforms all other proposed methods, both in terms of computing time and mixing.  The authors also provide a useful review of the algorithm and indicate its broad flexibility.  Throughout, we use the block Gibbs sampler for updating the matrix $\bK$ in the posterior.\\
\indent Both \citet{dobra_et_2011} and \citet{wang_li_2012} show that the random walk MH (RWMH) algorithm of \citet{mitsakakis_et_2011} is unsuitable for posterior inference.  However, we show below that it is especially effective to use in the double MH algorithm.  In particular, suppose that we wish to sample from  $\mathcal{W}_G(\delta, \mathbb{I}_p)$ and $\bK$ is the current state of an MCMC chain.  Then the RWMH algorithm performs the following
\begin{itemize}
\item[1.] Determine $\bPhi$ from $\bK$
\item[2.] Propose $\bPsi$:
\begin{itemize}
\item[a.] Sample $c\sim \chi^2_{\delta + \nu_i^G}$ and set $\Psi_{ii} = c^{1/2}$
\item[b.] Sample $\Psi_{ij} \sim \mc{N}(0,1)$ for $(i,j)\in G$
\item[c.] Complete $\bPsi$ using (\ref{eq:completephi}) for all $(i,j) \in F\setminus G, i < j$
\end{itemize}
\item[3.]  Compute
\begin{equation}\label{eq:rwmh_update}
\alpha = \exp\left(-\frac{1}{2}\sum_{(i,j)\in F\setminus G}(\Psi_{ij}^2 - \Phi_{ij}^2)\right)
\end{equation}
\item[4.] With probability $\min\{\alpha,1\}$ set $\bK = \bPsi'\bPsi$
\end{itemize}
The appeal of the RWMH algorithm in sampling from $\mc{W}_G(\delta,\mathbb{I}_p)$ is the simplicity of the factor in (\ref{eq:rwmh_update}).  Through the use of node reordering software, which minimizes the size of $F\setminus G$, this expression may require few calculations.  While the algorithm does not perform well in the posterior, and the calculation $(\ref{eq:rwmh_update})$ as well as step (2.c) become more involved when $\bD \neq \mathbb{I}_p$ we show below that in the particular case of the double MH algorithm using the prior $\mc{W}_G(\delta,\mathbb{I}_p)$, this method is extremely useful.
\subsection{Conditional Bayes Factors}\label{sec:cbfs}
\indent Prior to \citet{wang_li_2012}, model moves between two graphs $G$ and $G'$ focused on approximating the ratio
\begin{equation}\label{eq:bf}
\frac{pr(G|\mc{D})}{pr(G'|\mc{D})} = \frac{pr(\mc{D}|G)}{pr(\mc{D}|G')}\times \frac{pr(G)}{pr(G')},
\end{equation} 
first through MC \citep{atay-kayis_massam_2005,jones_et_2005}, then a combination of MC and Laplace approximation \citep{lenkoski_dobra_2011} and ultimately through RJ \citep{dobra_lenkoski_2011,dobra_et_2011}.\\
\indent Suppose that $G\subset G'$ which differ only by the edge $e = (i,j)\in G'$ and that $\bK\in\Pos_G$.  Let $\bK^{-e} = \bK\setminus\{K_{ij},K_{ji},K_{jj}\}$.  In lieu of (\ref{eq:bf}), \citet{wang_li_2012} consider ratios of the form
\begin{equation}\label{eq:cbf_k}
\frac{pr(G|\bK^{-e},\mc{D})}{pr(G'|\bK^{-e},\mc{D})} = \frac{pr(\mc{D},\bK^{-e}|G)}{pr(\mc{D},\bK^{-e}|G')}\times\frac{pr(G)}{pr(G')}
\end{equation}
which are related to the conditional Bayes factors (CBFs) of \cite{dickey_gunel_1978}.\\
\indent Using properties related to the form (\ref{eq:bgibbs_update}) \citet{wang_li_2012} show that 
\begin{equation}\label{eq:wang_li_cbf}
\frac{pr(\mc{D},\bK^{-e}|G)}{pr(\mc{D},\bK^{-e}|G')} = H(\delta + n,e,\bK^{-e},\bD^{*}) \frac{I_G(\delta,\bD)}{I_{G'}(\delta,\bD)}
\end{equation}
where, in general
$$
H(d,e,\bK^{-e},\bS) = \frac{I(d, S_{jj})}{J(d, S_{ee}, A_{11})}\left(\frac{|K^{0}_{V\setminus j}|}{|K^{1}_{V\setminus e}|}\right)^{(d - 2)/2}\exp\left(-\frac{1}{2}\inner{\bS, K^{0} - K^{1}}\right)
$$
where $I(b,c) = c^{-b/2}2^{b/2}\Gamma(b/2)$, 
$$
J(h,\bB,b) = \left(\frac{2\pi}{B_{22}}\right)^{1/2}b^{\frac{(h-1)}{2}}I(h,B_{22})\exp\left(-\frac{b}{2}\left[B_{11}-\frac{B_{12}^2}{B_{22}}\right]\right),
$$
such that $\bA = \bK_{ee} - \bK_{e,V\setminus e} \bK^{-1}_{V\setminus e} \bK_{e,V\setminus e}$.  The matrix $\bK^{0}$ is equal to $\bK$ except that $K^{0}_{jj} = \bK_{j,V\setminus j} \bK^{-1}_{V\setminus j} \bK_{j,V\setminus j}$ and $K^{0}_{ij}=K^{0}_{ji} = 0$.  Finally, the matrix $\bK^{1}$ is equal to $\bK$ except that $\bK^{1}_{e} = \bK_{e,V\setminus e} \bK^{-1}_{V\setminus e} \bK_{e,V\setminus e}$.\\
\indent By using the CBF in (\ref{eq:wang_li_cbf}), \citet{wang_li_2012} propose model moves that do not rely on RJ methods, and after assessing which graph to move to, the parameter $K_{jj}$, as well as $K_{ij}$ if $e$ is in the accepted graph, are resampled according to their conditional distributions given $\bK^{-e}$.  This method is appealing, as it offers an automatic manner of moving between graphs and does not rely on the tuning parameters used in the RJ methods of \citet{dobra_lenkoski_2011} and \citet{dobra_et_2011}.\\
\indent While the result has significant theoretical appeal we show that computation of the factor $H(\delta + n,e,\bK^{-e},\bD^{*})$ is extremely costly, even in low dimensions.  This is due to the formation of the matrices $\bK^{0}$ and $\bK^{1}$, which require the solution of systems involving large matrices, in particular, $\bK_{V\setminus j}$ and $\bK_{V\setminus e}$.\\\
\indent Suppose now that $G$ and $G'$ differ only by the edge $f = (p - 1,p)$ again with $G\subset G'$.  We consider the CBF
$$
\frac{pr(\mc{D},\bPhi^{-f}|G')}{pr(\mc{D},\bPhi^{-f}|G)},
$$
where $\bPhi^{-f} = \bPhi\setminus\{\Phi_{p-1,p},\Phi_{pp}\}$.  In Appendix A we show that
\begin{equation}\label{eq:cl_update}
\frac{pr(\mc{D}|\bPhi^{-f}, G')}{pr(\mc{D}|\bPhi^{-f},G)} = N(\bPhi^{-f},\bD^{*})\frac{I_{G}(\delta,D)}{I_{G'}(\delta,D)},
\end{equation}
with, in general
$$
N(\bPhi^{-f},\bS) = \Phi_{p-1,p-1}\left(\frac{2\pi}{S_{pp}}\right)^{1/2}\exp\left(\frac{1}{2} S_{pp} (\phi_0 - \mu )^2\right)
$$
where $\mu = \Phi_{p-1,p-1} S_{p-1,p}/S_{pp}$, and $\phi_0 = -\Phi_{p-1,p-1}^{-1}\sum_{l=1}^{p-2}\Phi_{lp-1}\Phi_{lp}$.\\
\indent This result originally appeared in an early version of \citet{wang_li_2012}.  In order to update a general edge $e$, we propose determining a permutation $\vartheta$ of $V$ such that the nodes of $V\setminus e$ are reordered to reduce the fill-in of the graph $G_{V\setminus e}$ and finally, the edge $e$ is placed in the $(p - 1,p)$ position.  Equation (\ref{eq:cl_update}) is then calculated after permuting, $\bK$ and $\bD^{*}$ according to $\vartheta$.\\
\indent The benefit of this method is the reduced computational overhead required to compute (\ref{eq:cl_update}).  The method requires relabeling the matrices $\bK$ and $\bD^{*}$ and determining the Cholesky decomposition of the permuted version of $\bK$.  Using node reordering software to minimize fill-in of $G_{V\setminus e}$ proves useful in the developments below.
\subsection{Avoiding Normalizing Constant Calculation}\label{sec:dmh}
\indent Both (\ref{eq:wang_li_cbf}) and (\ref{eq:cl_update}) require determination of the prior normalizing constants $I_G$ and $I_{G'}$.  While the MC method of \citet{atay-kayis_massam_2005} enables these factors to be approximated, the routine can be subject to numerical instability \citep{lenkoski_dobra_2011,wang_li_2012} and involves significant computational effort.\\
\indent \citet{wang_li_2012} propose a method for avoiding the use of the MC approximation for prior normalizing constants.  Their method employs the double Metropolis Hastings algorithm of \citet{liang_2010}, which is an extension of the exchange algorithm developed by \citet{murray_et_2006}.\\
\indent We briefly review the implementation of the double MH algorithm in \citet{wang_li_2012}.  Suppose that $(K,G)$ is the current state of the MCMC chain and we propose to move to $G'$ by adding the edge $e$ to $G$. The double MH algorithm then forms a copy $\tilde{\bK}$ of $\bK$, resamples $\tilde{K}_{ij}$ and $\tilde{K}_{jj}$ according to $G'$.  It then updates $\tilde{\bK}$ via block Gibbs according to $G'$.  Equation (\ref{eq:wang_li_cbf}) is then replaced with 
\begin{equation}\label{eq:wang_li_dmh}
\frac{H(\delta + n, e,\bK^{-e},\bD^{*})}{H(\delta, e,\tilde{\bK}^{-e},\bD)}
\end{equation}
We see that the expression (\ref{eq:wang_li_dmh}) has replaced the prior normalizing constants with an evaluation of $H$ in the prior, evaluated at $\tilde{K}$ \citep[see][for theoretical justifications of this procedure]{murray_et_2006,liang_2010}. This is clearly beneficial, as it avoids the need for involved MC approximation.  Unfortunately, the procedure as implemented in \citet{wang_li_2012} requires a full run of the block Gibbs sampler, as well as determination of the matrices $\bK^{0}$ and $\bK^{1}$ and therefore contains many large matrix operations.\\
\indent We propose an alternative implementation of the double MH algorithm.  Again suppose that $(\bK,G)$ is the current state and we propose to move to $G'$ by adding the edge $f = (p-1,p)$.  We first determine $\bPhi$ from $\bK$.  We then update $\bPhi$ to $\tilde{\bPhi}$ using the RWMH algorithm in Section~\ref{sec:gwish_samplers} relative to $G'$.  Equation (\ref{eq:cl_update}) is then replaced with
\begin{equation}\label{eq:cl_dmh}
\frac{N(\bPhi^{-f},\bD^{*})}{N(\tilde{\bPhi}^{-f},\bD)}
\end{equation}
\indent We can immediately see that the expression (\ref{eq:cl_dmh}) is considerably simpler that (\ref{eq:wang_li_dmh}); it requires no additional matrix inversions nor the evaluation of any trace inner products.   Furthermore, the generation of the auxiliary variables through the RWMH is considerably less demanding computationally than the use of the Block Gibbs sampler, especially when $\bD = \mathbb{I}_p$, a common setting in practice.
\subsection{Algorithms for Full Posterior Determination}\label{sec:algo}
In this section we outline the two algorithms we will consider for full posterior determination.  Both algorithms create a sequence $\{(\bK^{[1]},G^{[1]}),\dots, (\bK^{[S]},G^{[S]})\}$ where $\bK^{[s]} \in \Pos_{G^{[s]}}$.  Given the current state $(\bK^{[s]},G^{[s]})$ the WL algorithm proceeds as follows
\begin{itemize}
\item[0.] Set $\bK = K^{[s]}$ and $G = G^{[s]}$
\item[1.] For each edge $e$, do:
\begin{itemize}
\item[a.]  if $e \notin G$ attempt to update $G$ to $G' = G\cup e$ with probability
$$
\frac{q(G'|\bK^{-e},\mathcal{D})}{q(G|\bK^{-e},\mathcal{D})} = \frac{pr(G')H(\delta + n,e,\bK^{-e},D^{*})}{pr(G)}
$$
if $e \in G$ the ratio is flipped.  If $G$ is not to be updated, skip to step c. 
\item[b.]  If attempting to update $G$ to $G'$, sample $\tilde{\bK}$ as discussed in Section~\ref{sec:dmh} and calculate
$$
\alpha = \min\{1, H^{-1}(\delta,e,\tilde{\bK}, \bD)\}
$$
if $e\in G'$, otherwise calculate
$$
\alpha = \min\{1, H(\delta,e,\tilde{\bK}, \bD)\}
$$
and with probability $\alpha$ set $G = G'$, otherwise leave it unchanged.
\item[c.]  Resample $K_{ij}, K_{jj}$ according to $G$.
\end{itemize}
After attempting to update each edge, set $G^{[s + 1]} = G$.
\item[2.]  Resample $\bK^{[s + 1]}$ using the block Gibbs sampler relative to $G^{[s + 1]}$ and the current state of $\bK$.
\end{itemize}
We see that in one iteration of the WL algorithm, each edge is potentially updated in the graph.  Our new algorithm (which we call CL) also follows this idea, and proceeds as follows
\begin{itemize}
\item[0.] Set $\bK = \bK^{[s]}$ and $G = G^{[s]}$
\item[1.] For each edge $e$, do:
\begin{itemize}
\item[a.]  Determine a permutation $\vartheta$ of $V_p$ as discussed in Section~\ref{sec:cbfs}, which places the edge $e$ in the $(p-1,p) = f$ position, and likewise permute $\bK$, $G$, $\bD$ and $\bD^*$. 
Let $G^{\vartheta}$ denote the permuted version of $G$ and $\bPhi$ be the Cholesky decomposition of the permuted version of $\bK$.  If $f\notin G^{\vartheta}$ attempt to update $G^{\vartheta}$ to $G' = G^{\vartheta}\cup f$ with probability
$$
\frac{q(G'|\bPhi^{-f},\mathcal{D})}{q(G^{\vartheta}|\bPhi^{-f},\mathcal{D})} = \frac{pr(G')N(\bPhi^{-f},D^{*})}{pr(G^{\vartheta})}
$$
if $f\in G^{\vartheta}$ the ratio is flipped.  If $G^{\vartheta}$ is not to be updated then, skip to step c. 
\item[b.]  If attempting to update $G^{\vartheta}$ to $G'$, sample $\tilde{\bPhi}$ as discussed in Section~\ref{sec:dmh} and calculate
$$
\alpha = \min\{1, N^{-1}(\tilde{\bPhi}^{-f}, \bD)\}
$$
if $f\in G'$, otherwise calculate
$$
\alpha = \min\{1, N(\tilde{\bPhi}^{-f},\bD)\}
$$
and with probability $\alpha$ set $G^{\vartheta} = G'$, otherwise leave it unchanged.
\item[c.]  Resample $\Phi_{p-1,p}, \Phi_{pp}$ according to $G^{\vartheta}$.  Then reform $\bK$ and $G$ by unpermuting the system.
\end{itemize}
After attempting to update each edge, set $G^{[s + 1]} = G$.
\item[2.]  Resample $\bK^{[s + 1]}$ using the block Gibbs sampler relative to $G^{[s + 1]}$ and the current state of $\bK$.
\end{itemize}
As we can see, there is somewhat more bookkeeping involved in the implementation of the CL algorithm, as the system is constantly being permuted.  However, the reduction in computation time by using the RWMH algorithm and requiring only the calculation of the factors $N(\bPhi^{-f},\bD^{*})$ and $N(\tilde{\bPhi}^{-f},\bD)$ is dramatic, as we show below.\\
\section{Simulation Study}\label{sec:simulation}
In this section we conduct a simulation study that compares the method we have developed to the WL algorithm. Our example comes directly from \citet{wang_li_2012}.  We consider a situation in which $p = 6$ and let $\bU = \bY\bY' = nA^{-1}$ where $n = 18$ and $A_{ii} = 1$ for $i = 1,\dots, 6; A_{i,i + 1} = A_{i + 1, i} = .5$ for $i = 1,\dots,5$ and $A_{16} = A_{61} = .4$.  We finally assume the prior $\bK \sim \mc{W}_G(3,\mathbb{I}_6)$.  Using exhaustive MC approximation of the entire graph space, \citet{wang_li_2012} show that the posterior probability of each edge is
$$
(p_{ij}|A) = \left(\begin{array}{cccccc} 1 & 0.969 & 0.106 & 0.085 & 0.113 & 0.85\\
0.969 & 1 & 0.98 & 0.098 & 0.081 & 0.115\\
0.106 & 0.98 & 1 & 0.982 & 0.0098 & 0.086\\
0.085 & 0.098 & 0.982 & 1 & 0.98 & 0.106\\
0.113 & 0.081 & 0.098 & 0.98 & 1 & 0.97\\
0.85 & 0.115 & 0.086 & 0.106 & 0.97 & 1\\
\end{array}
\right)
$$
\indent We use this example and compare the CL algorithm to the WL algorithm.  Following \citet{wang_li_2012} we run both the WL and CL algorithms as described in Section~\ref{sec:algo} for $60,000$ iterations and discard the first $10,000$ iterations as burn-in.  Both algorithms were implemented in {\tt R}, though {\tt C++} was used for block-Gibbs updates.  We note that if a pure {\tt R} implementation had been used, the time differences between WL and CL would be even more dramatic.\\
\indent We record the total computing time and looked at the mean squared errors of the posterior inclusion probabilities from the two runs compared with the true values given above.  We repeated the entire process $100$ times, each time starting both WL and CL from the same random starting point.  Table~\ref{tbl:wl_sim} shows the average computing time in seconds (on a 2.8 GHz desktop computer with 4GB of RAM running Linux), average MSE and standard deviations across the $100$ runs.  The first column shows the expected result: even in six dimensions the WL algorithm takes more than 4 times as long to perform the same number of iterations as the CL algorithm.  This shows the improved efficiency of the proposed method.\\
\begin{table}
\caption{Comparison of CL and WL algorithms for the six dimensional example.}\label{tbl:wl_sim}
\begin{center}
\begin{tabular}{lcccc}
\hline\hline
&\multicolumn{2}{c}{Time (sec)}&\multicolumn{2}{c}{MSE}\\
&Mean&SD&Mean&SD\\
\hline
CL & 182.5 & (4.1) &0.0088& (6e-04)\\
WL & 818.4 & (19.2) &0.0349& (0.0025)\\
\hline\hline
\end{tabular}
\end{center}
\end{table}
\indent We found the results in the third column surprising, but do not draw broad conclusions from it.  It appears that in this example, using $60,000$ iterations, the CL algorithm approaches the true posterior edge expectation more quickly than the WL algorithm.  Since both algorithms are correct theoretically, we choose not to emphasize this result.  Furthermore, we have determined that by doubling the number of iterations, both approaches yield essentially the exact posterior distribution, though again the WL algorithm takes more than 4 times as long to run.\\
\indent This example was chosen as it appears in \citet{wang_li_2012} and has an exact answer.  The fact that the CL algorithm is considerably faster than the WL approach even in 6-dimensions indicates the broader appeal for searching truly high dimensional spaces.
\section{A Multivariate Graphical Stochastic Volatility Model}\label{sec:vol}
\indent Modeling the joint distribution of returns for a large number of assets is an important component of portfolio allocation and risk management.  \citet{carvalho_west_2007} and \citet{rodriguez_et_2011} both show that the use of GGMs can substantially improve modeling of joint asset returns.  However, heterogeneity in asset returns was, at best, tangentially addressed.  The study of \citet{carvalho_west_2007} considered a fixed (decomposable) graph throughout the entire period and assumed that asset returns were identically distributed.  \citet{rodriguez_et_2011} allowed for mixing over the class of decomposable graphs and also introduced some heterosckedasticity by considering an infinite-dimensional hidden Markov model (iHMM).  However, inside groups of observations in the iHMM, variances were assumed constant.\\
\indent Despite these constraints in model assumptions, both studies showed substantial improvements by incorporating GGMs into the precision matrix associated with joint asset returns.  We consider a situation in which the notion of homoskedastic, normally distributed asset returns is simply untenable; namely, the period surrounding the financial turbulence associated with the collapse of Lehman Brothers.  We show that by utilizing the developments in the previous sections, we are able to specify a parsimonious stochastic volatility model for multivariate assset returns that quickly adapts to changes in market volatility.  This model shows the flexibility of the new approach in embedding the GGM in larger hierarchical Bayesian frameworks.
\subsection{The stochastic volatility model}\label{sec:model}
\indent Let $\bY_t$ be the log-returns of $p$ correlated assets.  We specify the following hierarchical model for these returns
\begin{align}
\bY_{t}|\bK,X_t &\sim \mc{N}_p(\bs{0}, \exp(X_t) \bK^{-1})\label{eq:stoch_vol_lik}\\
X_t|\phi,X_{t - 1},\tau &\sim \mathcal{N}(\phi X_{t -1}, \tau^{-1}) \nonumber\\
\phi &\sim \mc{N}(0,\tau_0)\nonumber\\
\tau &\sim \Gamma(a,b)\nonumber\\
K|G &\sim \mc{W}_{G}(\delta,\bD)\nonumber\\
G &\sim pr(G)\nonumber.
\end{align}
\indent In the likelihood (\ref{eq:stoch_vol_lik}) we see that asset returns are assumed to be mean-zero.  The $X_t$ terms then dictate an overall level of market volatility, while a constant precision parameter $\bK$ dictates the degree to which asset returns are correlated.  While this model is parsimonious, it serves as a useful first departure from previous studies as it explicitly incorporates notions of stochastic volatility.  For purposes of identification, we set $X_0 = 0$.  In the conclusions section we discuss further possible generalizations to this framework.\\
\indent After collecting a time-series of returns $\bY^{(1:T)}$, we then aim to determine the posterior distribution
$$
pr(K,G,\tau,\bs{\phi},\bX|\bY^{(1:T)})
$$
where $\bX = (X_1,\dots,X_T)$.  Furthermore, we may be interested in the posterior predictive distribution $pr(\bY^{(T + 1)}|\bY^{(1:T)})$.  The parameters $\bX, \phi, \tau$ are updated with standard block MH or Gibbs steps \citep[see][]{rue_held_2005} and the posterior predictive distribution is easily formed from these parameters.  However, we note in particular that
\begin{equation}
\bK|G,\tau,\bs{\phi},\bX,\bY^{(1:T)}\sim \mc{W}_G\left(\delta + T, \bD +\sum_{t = 1}^{T} \frac{\bY^{(t)}\bY^{(t)'}}{\exp(X_t)}\right)\label{eq:stoch_post}.
\end{equation}
From (\ref{eq:stoch_post}) we see why the developments in Section~\ref{sec:gwish} prove useful.  We may update $\bK$ and $G$ jointly using the CL algorithm discussed in Section~\ref{sec:algo} simply by setting $\bD^{*} = \bD + \sum_{t = 1}^{T} (\bY^{(t)}\bY^{(t)'})/\exp(X_t)$.  This allows us to easily embed a sparse precision matrix $\bK$ and mix over the class of GGMs in any hierarhical Bayesian model that involves a standard Wishart distribution.
\subsection{The data}\label{sec:data}
\begin{figure}
\centering
  \includegraphics[width=0.5\linewidth,trim = 0cm 1cm
  0cm 2cm]{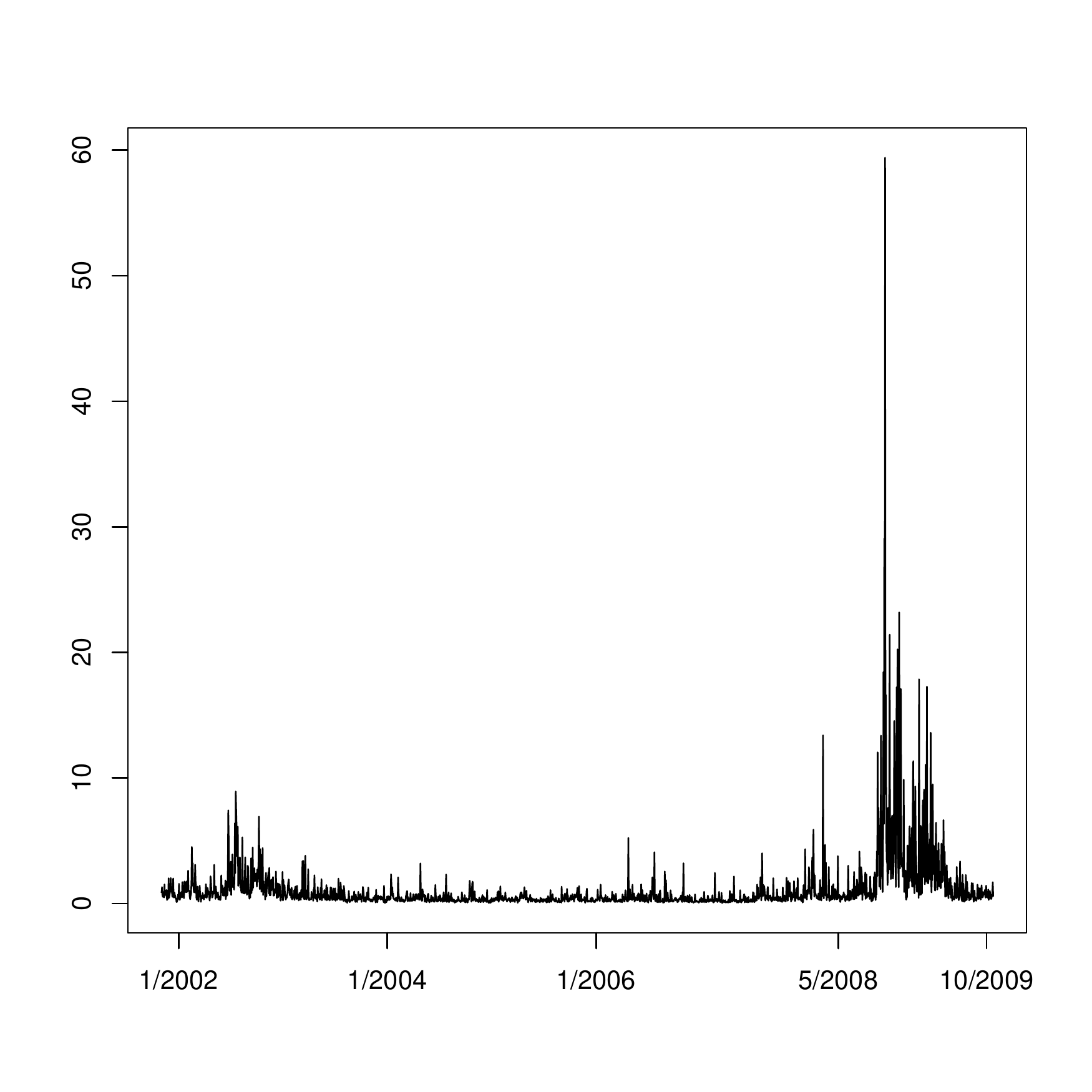}
  \caption{Mean of the squared returns taken over all 20 stocks during
    the entire time period from October 31, 2001 to October 23, 2009.}\label{fig:square_returns}
\end{figure}

To apply our model and algorithm we randomly chose 20 stocks from the
S\&P 500. These stocks were: Aetna Inc. (AET), CA Inc. (CA), Campbell Soup (CPB), CVS Caremark Corp. (CVS), Family Dollar Stores (FDO), Honeywell Int'l Inc. (HON),
Hudson City Bancorp (HCBK), JDS Uniphase Corp. (JDSU), Johnson Controls (JCI), Morgan Stanley (MS), PPG Industries (PPG),
Principal Financial Group (PFG), Sara Lee Corp. (SLE), Sempra Energy (SRE), Southern Co. (SO), Supervalu Inc. (SVU), Thermo Fisher Scientific (TMO),
Wal-Mart Stores (WMT), Walt Disney Co. (DIS), Wellpoint Inc. (WLP).\\
\indent We chose a time period where markets experience both high and low volatility to evaluate the flexibility of our model. We chose the time
period from October 31, 2001 to May 21, 2008 as our training period to
fit our model and make predictions for the time period from May
22, 2008 to October 23, 2009. The time periods consist of 1650 and 360
trading days respectively. Figure~\ref{fig:square_returns} shows the mean of the squared returns for the these 20 securities over the entire dataset.  The extreme volatility present in the markets after the collapse of Lehman brothers in September 2008 is readily evident, showing that a homoskedasticity assumption is untenable for these data.
\subsection{Predictive Performance Results}
\begin{figure}
  \centering
  \includegraphics[width=0.6\linewidth,trim= 0cm 1.5cm
  0cm 1cm]{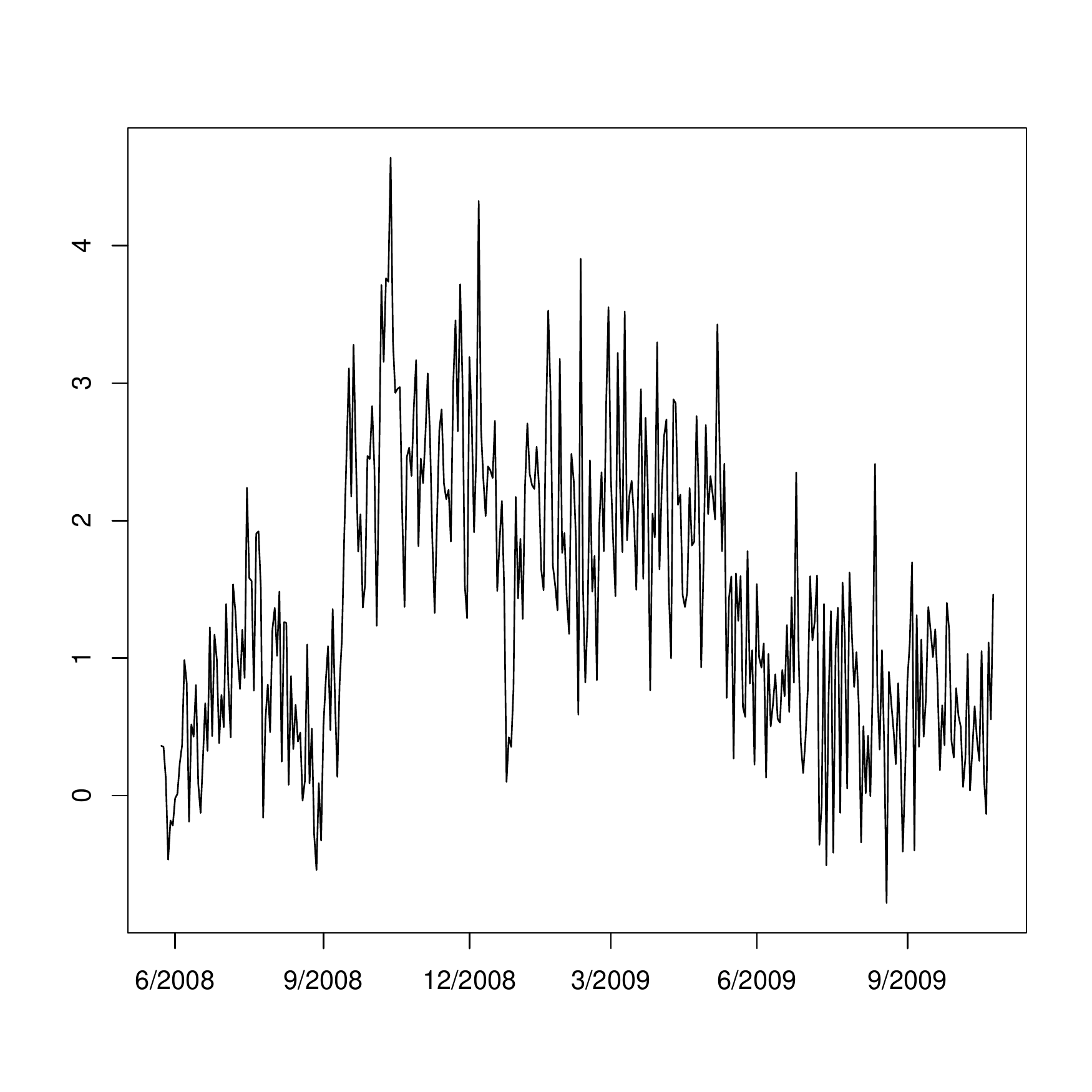}
    \caption{Means of posterior predictive distribution for the volatility component $X_{t +1}$ from May 22,
      2008 to October 23, 2009.}\label{fig:vol_means}
\end{figure}

\begin{figure}
  \centering
  \includegraphics[width=0.6\linewidth,trim= 0cm 1.5cm
  0cm 1cm]{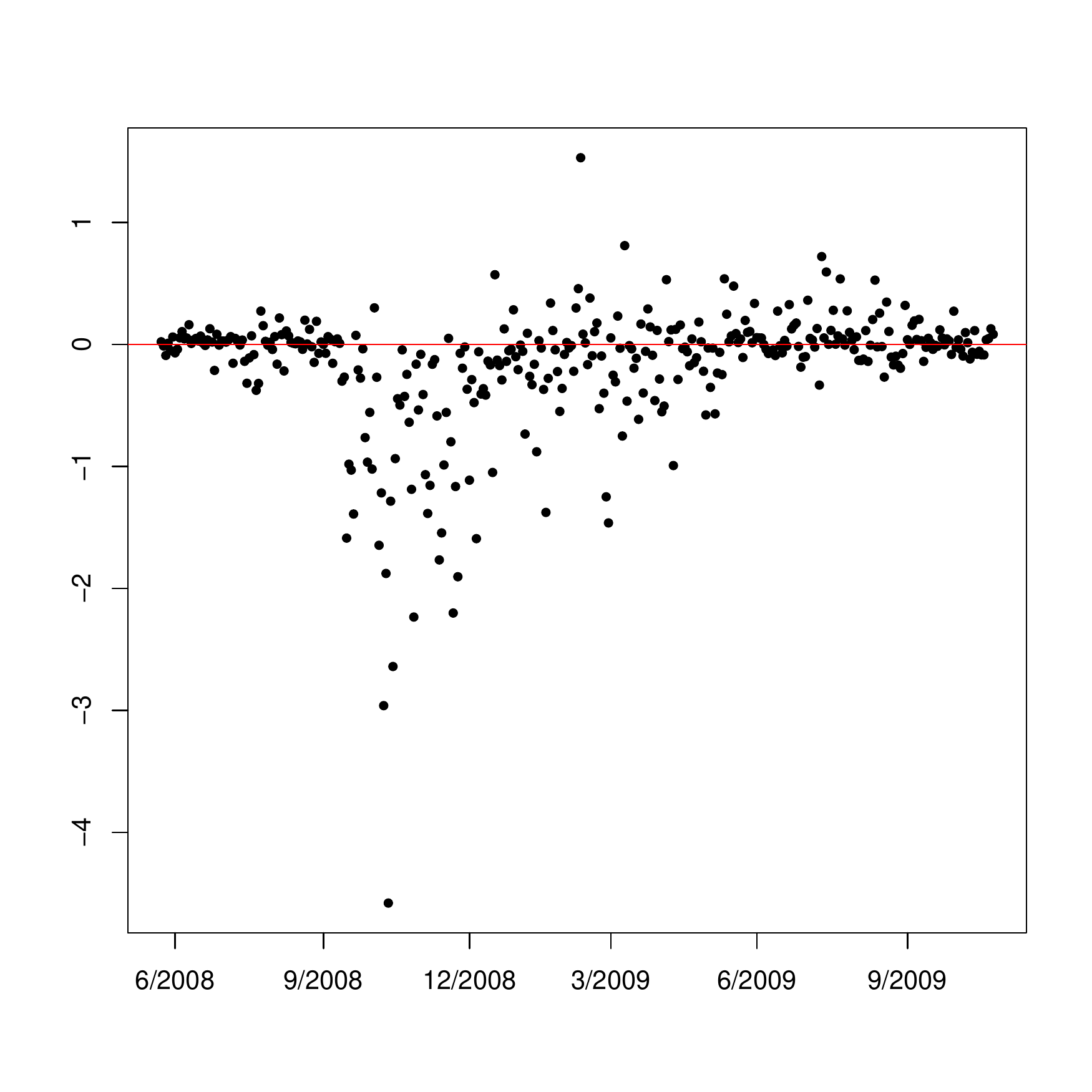}
    \caption{Difference of energy score from the predictive distribution of the model with stochastic volatility versus the model with fixed volatility.  Values below zero indicate the stochastic volatility model outperformed the fixed volatility model.}\label{fig:es_diff}
\end{figure}

We assess the relative performance of the stochastic volatility model we develop in Section~\ref{sec:model} versus a method that embedds GGMs, but does not have a stochastic volatility component. For each day $t+1$ in the forecast period we run our stochastic
volatility model from the beginning of the training period
until the previous day
$t$ to obtain estimates for the model parameters. We run the algorithm for $60,000$ iterations, discarding the first $10,000$ as burn-in (keep in mind that in one iteration of the algorithm all edges are evaluated).  Using the posterior sample, we obtain the posterior predictive distribution of $\bY^{(t + 1)}$.\\
\indent Figure~\ref{fig:vol_means} shows the mean of the posterior predictive distribution of the volatility component $X_{t + 1}$ using the returns up to time $t$, which drives the predictive distribution of $\bY^{(t + 1)}$ for each day in the forecast period.  Comparing Figures~\ref{fig:square_returns} and \ref{fig:vol_means} we see
that our model reflects the time-dependent volatility well. At the beginning when the market is quiet, $X_{t + 1}$ takes lower values mostly between 0 and 1. After the shock of the financial crisis the
volatility in the market goes up extremely, which is reflected by significantly higher values of $X_{t + 1}$. Months later, towards the end of the forecast period, the market has cooled down and the terms $X_{t + 1}$ reflect this.\\
The two methods we compare both return full predictive distributions.  By construction, these predictive distributions have the same mean and median since returns are assumed mean-zero.  Judging their performance therefore requires assessing the entire predictive distribution. We assess their performance using the energy score introduced by \cite{gneiting_raftery_2007}. \\
\indent
The energy score is a proper scoring rule, which is a multivariate
generalization of the continuous ranked probability score. It is defined as 
\begin{equation}
\label{eq:69}
  ES(F,\mathbf{x})= \mathbf{E}_F||\mathbf{X}-\mathbf{x}||^{2} - \frac12 \mathbf{E}_F ||\mathbf{X}-\mathbf{X}'||^{2}
\end{equation}
where $F$ is our predictive distribution for a vector-valued quantity,
$\mathbf{X}$ and $\mathbf{X}'$ are independent random variables with
distribution $F$ and
$\mathbf{x}$ is the realization.\\
\indent
For each day in the forecast period, we compute the energy score for the predictive distribution returned by the two methods considered.  Figure~\ref{fig:es_diff} shows the difference between the stochastic volatility model developed in Section~\ref{sec:model} and the model that incorporates GGM uncertainty but holds volatility fixed.  As we can see in Figure~\ref{fig:es_diff}, between May and August, 2008, there is no clear difference between the two approaches.  However, after the financial turbulence in September, 2008, the stochastic volatility model outperforms the fixed volatility model by a considerable margin.  During almost every day in the turbulent period, the energy scores are lower under the stochastic volatility model.  After the market turbulence subsides, the two models return to performing equally well again.\\
\indent This short example shows the utility of the computational methodology developed in this paper.  The model is simple, in many respects, but a non-trivial deviation from the standard iid sampling framework to which the GGM was initially relegated.  By now being able to embed the GGM in more complicated hierarchical frameworks, we are able to address difficult sampling schemes while simultaneously incorporating sparsity in the estimate of joint distributions.

\section{Conclusions}\label{sec:conclude}
\indent We have synthesized a number of recent results related to the G-Wishart distribution.  This has allowed for an algorithm that does not rely on RJ methods, obviates the need for expensive and numerically unstable MC approximation of prior normalizing constants and does so with minimal computational effort.  The improvement in computation time is sufficient that at each stage of the algorithm, all edges may be evaluated for inclusion or exclusion in the graphical model.  This algorithm allows the GGM to be embedded in more sophisticated hierarchical Bayesian models and opens the possibility of replacing standard Wishart distributions with G-Wishart variates, leveraging the improvement in predictive performance offered by sparse precision matrices.\\
\indent The applied example shows the usefulness of this combination.  We are able to sparsely model the interactions in financial assets while simultaneously addressing the issues of stochastic volatility prevalent in markets undergoing turbulence.  The method is able to characterize the distribution of asset returns during periods of rapidly fluctuating volatility much better than standard iid frameworks.\\
\indent The stochastic volatility model we develop remains parsimonious and several adjustments could be made.  The first such development would be to replace the univariate term $X_t$ with a multivariate factor that allows the variance of each asset to follow its own path, while potentially tying the evolution of these factors together with a separate GGM.  Furthermore, employing some form of the iHMM framework of \citet{rodriguez_et_2011} could allow for the matrix $\bK$ to change throughout the period as well.  Such developments will be considered in future work.\\
\bibliographystyle{apalike}
\bibliography{ChengLenkoski}

\section*{Appendix A: Determination of CBF for using $\bPhi^{-f}$}
Consider
$$
\frac{pr(\mc{D},\bPhi^{-f}|G')}{pr(\mc{D},\bPhi^{-f}|G)}
$$
we note that
$$
pr(\mc{D},\bPhi^{-f}|G') = \int_{\Phi_{p-1,p}}\int_{\Phi_{pp}} pr(\mc{D},\bPhi|G')d\bPhi_{f}
$$
and
$$
pr(\mc{D},\bPhi^{-f}|G) = \int_{\Phi_{pp}} pr(\mc{D},\bPhi|G)d\Phi_{pp}
$$
up to common terms we thus have that
$$
pr(\mc{D},\bPhi^{-f}|G') \propto \frac{\Phi_{p-1,p-1}}{I_{G'}(\delta,\bD)}\int_{\Phi_{p-1,p}} \exp\left(-\frac{1}{2} D^{*}_{p,p}(\Phi_{p - 1,p} + \mu)^2\right)d\Phi_{p-1,p}
$$
recognizing the integral as the kernel of a normal distribution, this yields
$$
pr(\mc{D},\bPhi^{-f}|G') \propto \frac{\Phi_{p-1,p-1}}{I_{G'}(\delta,\bD)}\left(\frac{2\pi}{D^{*}_{pp}}\right)^{1/2}.
$$
Further, again up to common terms
$$
pr(\mc{D},\bPhi^{-f}|G) \propto \frac{1}{I_{G}(\delta,\bD)}\exp\left(-\frac{1}{2} D^{*}_{p,p}(\Phi_{0} + \mu)^2\right)
$$
and thus
$$
\frac{pr(\mc{D},\bPhi^{-f}|G')}{pr(\mc{D},\bPhi^{-f}|G)} = N(\bPhi^{-f},D^{*})\frac{I_G(\delta,\bD)}{I_{G'}(\delta,\bD)}
$$
\end{document}